\documentclass[twoside,11pt]{article}

\usepackage{jmlr2e}
\usepackage{amssymb}
\usepackage[english]{babel}
\usepackage{amsfonts}
\usepackage{makeidx}
\usepackage{hyperref}
\usepackage{multirow}
\usepackage{caption}
\usepackage{amssymb,amsmath,amsfonts}
\usepackage[left,modulo]{lineno}
\usepackage{bm}
\usepackage{psfrag}
\usepackage{graphics}
\usepackage{fancyhdr}
\usepackage{eucal}
\usepackage[english]{babel}
\usepackage{ifpdf}
\usepackage{makeidx}
\usepackage{setspace}
\usepackage{float}
\usepackage[T1]{fontenc}
\usepackage{subfigure}
\usepackage{dsfont}
\usepackage[ruled, vlined, linesnumbered, longend]{algorithm2e}
\usepackage{graphicx, framed}
\usepackage{textcomp}

\usepackage{epstopdf}

\usepackage{tikz}
\usepackage{tikz-inet}
\usetikzlibrary{shapes,snakes,arrows,automata}

\newif\ifnotes\notestrue

\def\hgr#1{}

\begin{document}

\title{Hidden Markov Models for Gene Sequence Classification: Classifying the VSG genes in the Trypanosoma brucei Genome}

\author{
\name Andrea Mesa\email andreagmesa@ccee.edu.uy\\
\addr Facultad de Ciencias Econ\'omicas y Administraci\'on, \\
Universidad de la Rep\'ublica,\\
Montevideo 11200, Uruguay\\
\AND
\name Sebasti\'an Basterrech\email Sebastian.Basterrech.Tiscordio@vsb.cz\\
       \addr National Supercomputer Center\\
       V\v{S}B-Technical University of Ostrava\\
       Ostrava, 708 00, Czech Republic
\AND
\name Gustavo Guerberoff\email gguerber@fing.edu.uy\\
\addr Instituto de Matem\'atica y Estad\'istica, Facultad de Ingenier\'ia\\
Universidad de la Rep\'ublica,\\
Montevideo 11200, Uruguay\\
\AND
Fernando Alvarez-Valin\email falvarez@fcien.edu.uy\\
\addr Sección Biomatem\'atica - Facultad de Ciencias\\
Universidad de la Rep\'ublica,\\
Montevideo 11400, Uruguay\\
}

\editor{--}

\maketitle

\begin{abstract}
The article presents an application of \textit{Hidden Markov Models (HMMs)} for pattern recognition on genome sequences. We apply HMM for identifying genes encoding the \textit{Variant Surface Glycoprotein (VSG)} in the genomes of \textit{Trypanosoma brucei (T. brucei)} and other African trypanosomes. These are parasitic protozoa causative agents of sleeping sickness and several diseases in domestic and wild animals. These parasites have a peculiar strategy to evade the host's immune system that consists in periodically changing their predominant cellular surface protein (VSG). The  motivation for using patterns recognition methods to identify these genes, instead of traditional homology based ones, is that the levels of sequence identity (amino acid and DNA sequence) amongst  these genes is often below of what is considered reliable in these methods.  Among pattern recognition approaches, HMM are particularly suitable to tackle this problem because they can handle more naturally the determination of gene edges. We evaluate the performance of the model using different number of states in the Markov model, as well as several performance metrics. The model is applied using public genomic data. Our empirical results show that the VSG genes on \textit{T. brucei}  can be safely identified (high sensitivity and low rate of false positives) using HMM.
\end{abstract}


\newcommand{\trans}{\bm{\varrho}}
\newcommand{\emis}{\bm{\varepsilon}}
\newcommand{\colec}{\bm{\theta}}

\section{Introduction}
\label{intro}
Nowadays, the community disposes of a huge amount of genomic sequential data.
Machine learning techniques are indispensable in the analysis of sequences due to the enormous size of the available genome databases.
In this article we apply \textit{Hidden Markov Models (HMMs)} for identifying a specific pattern in a hyper-variable gene sequences.
Markov models have been widely applied for modelling several complex systems that evolve in time.
In particular the HMM have proven to be a powerful tool for classification tasks on time series data and spatial sequential data.
During the last 20 years, the model has been used for analysing biological sequences, and for solving problems like gene finding and protein analysis.
We apply a HMM for identifying genes encoding the \textit{Variant Surface Glycoprotein (VSG)} in the genome of \textit{Trypanosoma brucei (T. brucei)}.
This approach can be naturally be extended to other African trypanosomes that also have this type of of hyper-variable genes.
These are parasitic protozoa causative agent of \textit{sleeping sickness} in humans and several diseases in domestic and wild animals. 
These parasites have a peculiar strategy to evade the host's immune system that consists in periodically changing their predominant cellular surface protein (VSG).

There are several methods that have been applied for detecting coding regions in a genome, among which the most used are those based on sequence comparison and \textit{ab initio} methods~\citep{Wang04}.
Homology gene finding methods are based in given a DNA sequence identifying its homologous sequences in other genomes stored in databases.
In \textit{ab initio} gene prediction the goal is finding either the nearby presence or statistical properties of the coding region i.e. use gene structure as a template to detect genes.
%
%
Although, for the particular case of the VSG genes these techniques sometimes do not yield reliable results, due to the high variability of the VSG genes.
It means that, the levels of sequence identity (amino acid and DNA sequence) amongst the VSG genes is often below of what is considered reliable in the traditional homology based methods.  Therefore in this article, we study the possibility of locating these genes looking for a statistical signature using HMM.
In the following, we begin by specifying the motivations for studying this particular genoma. Next, we clarify the contributions and goals of this article.

\subsection{Motivation}

The \textit{African Trypanosomiasis} also called \textit{sleeping sickness} is caused by  \textit{T. brucei} parasite.
There are two subspecies that causes disease in humans: \textit{T. brucei gambiense} and \textit{T. brucei rhodesiense}. A third sub-species named \textit{T. brucei brucei} exists although it rarely infects humans. 
Around the $90\%$ of the sleeping sickness is caused by \textit{T. brucei} gambiense~\citep{OMS}.
%
%
The disease infections occur in regions of the sub-Saharan Africa. 
The disease is propagated by the bite of an infected tsetse fly that acts as disease vector.
%
%
Even though over the last years the number of cases have decreased, in~$2009$ were reported  ~$9878$ cases~\citep{OMS}.
This decreasing tendency has continued, and it has been reported~$7216$ cases in~$2012$. 
%
%
%
%

%
%

Cell surface of the trypanosoma consists of a uniform layer of \textit{Variant Surface Glycoprotein (VSG)} that blocks the trypanosomes recognition by the immune system of the mammal host. 
Once the individual is infected, the trypanosoma expresses a particular VSG and the host immune system reacts by generating a response to this protein layer, which causes a decrease in the number of trypanosomes. 
Some trypanosomes can evade the recognition by the immune system. 
By expressing an alternative VSG for which have not yet developed antibodies. 
This allows increasing the population of infectious agents  sustaining a long term infection, thus increasing the chances of transmission. 
Therefore, various infections occur each of which is caused by cells expressing a different VSG variant.

\subsection{Objectives and Contributions}

HMM is a probabilistic model used for the analysis of sequential data.
There is a general consensus about the power of HMM for classifying time-series data and sequential data.
The goal is developing a machine learning tool based on HMM that identifies homogeneous zones with the VSG genes in the genome of \textit{T. brucei}.
Among pattern recognition approaches, HMM are particularly suitable to tackle this problem because they can handle more naturally the determination of gene edges.
As far as our knowledge is concerned, this article presents the first mathematical approach based on HMM for modelling such sequence.

The main contribution is proposing an algorithm for classifying zones of a genome on two classes, which is based on the following three features: gene location, correlation between symbols in nucleotide sequence and its hidden information included in that sequence.
%
%
%
%
The proposed approach is an hybrid procedure that combines the gene location and the power of the HMM technique for modelling the underlying structure of the sequence.
We analyse the performance of our procedure on a public data from \textit{GenBank} database~\citep{Benson09}. The performance of the algorithm is evaluated using different measures from the confusion matrix.
Despite its simplicity, according to our experimental results the algorithm is robust and performant in accuracy (high sensitivity and low rate of false positives) and time for detecting VSG genes on \textit{T. brucei}. 

\subsection{Organization}

This article is organized as follows. Next section is a general introduction to HMMs and its applications.
Section~\ref{Training} introduces the algorithms used for estimating the parameters of the model. 
In~\ref{BW} we present the Baum-Welch Algorithm, and in~\ref{Viterbi} we introduce the Viterbi algorithm. We present the methodology used for solving our problem in~\ref{Methodology}. 
The experimental results are presented in~\ref{Result}. 
Finally, we conclude the article and discuss future research lines.

\section{Hidden Markov Model Description}
\label{HMM}
In this Section, we describe in brief the \textit{Hidden Markov Model (HMM)}. 
We discuss the general assumptions and properties of a HMM, and we present the approach to use it on the gene identification problem. The end of this section presents some related works that have used HMM for classification on sequential data.
\subsection{General Introduction}
Markov models are one of the most powerful tools for stochastic modelling of dynamic systems. 
The theory of Markov is rich and quite complete, and the associated algorithmic tools are very efficient.
The Markov models are characterised for their high ability to capture all kinds of behaviors.
The model provides time-scale invariability in recognition and learning capabilities~\citep{Yamato92}.
%
%
A HMM is a particular time-discrete Markov model where the states are not directly visible.
%
The theory of HMM was developed by Baum in the late 60s~\citep{Baum67, Baum70}, and in 1989, Rabiner~\citep{Rabiner89} publishes a tutorial describing Baum theory applied to speech recognition that is a reference in the study of such models.
The family of HMMs has also shown to be effective for analysing biological sequence data~\citep{Yoon09,Durbin98}. The applications include: sequence alignment~\citep{Pachter02}, gene and protein structure predictions~\citep{Munch06,Won07}, modelling DNA sequencing errors~\citep{Lottaz03}, and for analysing RNA structure~\citep{Yoon08,Harmanci07}.  
A HMM is composed by the following elements~\citep{Rabiner86}: a finite set of states, a finite alphabet, and probabilities of state transition and symbol emission.
At each time $t$ a new state is entered in the system following a transition probability distribution that verifies the Markovian property, after each transition an output symbol is produced according to some probability distribution that depends on the current state.

Let $\cal A$ be a finite alphabet and $\cal E$ be a finite set of states or labels.
Given a sequence of observations $x=(x_1, x_2,\ldots,x_n)$, $x_j\in\cal A$ and a collection of states $y=(y_1, y_2, \ldots, y_n)$, $y_j\in\cal E$ for all $j$, where each $x_j$ has been emitted by its state $y_j$.
A HMM describes how the sequence $x$ and its associated state $y$ progresses in time. Note that, the index $j$ ($j=1,\ldots, n$) can be thought of as a time index.
We say that the states are \textit{hidden} due to the sequence of observations is known but its related states are unknown.
%
%
The model works as follows: it starts in an initial state chosen according to some probability, it emits an observation, moves to a new state emits a new observation and so on until to reach some final conditions.

\begin{figure}[h]
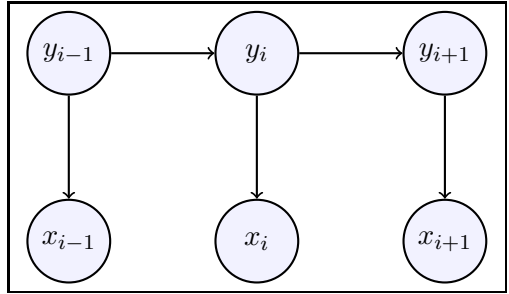

\begin{center}
\fbox{
	\tikz{
		\node[draw,thick,fill=blue!5, circle,minimum size=11mm]  (xi-1) at (0,0) {$x_{i-1}$};
		\node[draw,thick,fill=blue!5, circle,minimum size=11mm]  (xi) at (2.5,0) {$x_{i}$};
		\node[draw,thick,fill=blue!5, circle,minimum size=11mm]  (xi+1) at (5,0) {$x_{i+1}$};
		\node[draw,thick,fill=blue!5, circle,minimum size=11mm]  (yi-1) at (0,2.5) {$y_{i-1}$};
		\node[draw,thick,fill=blue!5, circle,minimum size=11mm]  (yi) at (2.5,2.5) {$y_{i}$};
		\node[draw,thick,fill=blue!5, circle,minimum size=11mm]  (yi+1) at (5,2.5) {$y_{i+1}$};
		\draw[->,black] (yi-1) edge[->,thick] (yi);
		\draw[->,black] (yi) edge[->,thick] (yi+1);
		\draw[->,black] (yi-1) edge[->,thick] (xi-1);
		\draw[->,black] (yi) edge[->,thick] (xi);
		\draw[->,black] (yi+1) edge[->,thick] (xi+1);
		}
	}
\caption{\label{RepresentationHMM}Usual representation of a HMM. Figure illustrates the conditional independence relationship among the states and the observations in the system.}
\end{center}
\end{figure}

The model parameters are the probability of emitting symbols and the probabilities of the state transitions.
Let $\mathbb {P}\left(x_i\vert y_i=k \right)$ be the emission probability that the symbol $x_i$ is generated when the sequence is at the state $k$ at time $i$.
%
%
The probability of transition from the state $k$ to state $l$ is given by $\displaystyle{\mathbb {P} \big(y_{i+1}=l|y_{i}=k\big)}$, i.e. the conditional probability that the sequence is in the state $l$ at time $i+1$ when at time $i$ was in the state $k$.
For simplicity the above probabilities are respectively written by $\mathbb{P}\left(y_{i+1}\vert y_{i} \right)$ and $\mathbb{P}\left(x_i\vert y_i \right)$.
At $n$ steps the model generates a sequence $x=(x_1, x_2, \ldots, x_n)$ for which passes through a sequence of states $y=(y_1, y_2, \ldots, y_n)$. 
For a fixed length $n$ the model defines a probability distribution over all sequences of observations $x$ and all possible states $y$. 
Then, the probability that the model transits the path $y$ and generates the sequence $x$ is given by
\begin{equation}
\mathbb{P}(x, y)=\prod_{i=1}^{n} \mathbb{P}\left(y_{i}\vert y_{i-1} \right)\mathbb{P}\left(x_i \vert y_i \right),
\end{equation}
where the first transition $\mathbb{P}(y_1\vert y_0)$ is given by $\mathbb{P}(y_1)=\mathbb{P}(y_1\vert y_0)$.
Figure~\ref{RepresentationHMM} illustrates the relationship among states and observations in a HMM. We can see the conditional independence among the events in the system. The probability of the state $y_{i}$ only depends on the previous state $y_{i-1}$, and the observation generated at time $i$ only depends on the state at time $y_i$.

In this article, we denote the unknown model parameters using greek letters, the matrix of transition probabilities by $\trans$, the emission matrix by $\emis$, and the collection of the whole parameter of the model by $\bm{\colec}$.
We use bold fonts for matrices and vectors and normal fonts for scalars.
Those parameters are estimated in a training step by a particular case of the Expectation Maximization algorithm named the Baum-Welch algorithm~\citep{Welch03Lecture,LibroHastie}.
%
%
%
The model is based in the following assumptions:
\begin{itemize}
\item Markov assumption: the transition probability among the states is defined by $\mathbb{P} \big (y_ {i+1}| y_i \big)$. This means that the next state depends only on the current state that is called Markov assumption of first order.
%

\item Homogeneity: it is assumed that the state transition probabilities are homogeneous, i.e. they are independent of its location in the sequence. 
As a consequence, the probability of moving from one state to the next state is independent on the position of the sequence in which this transition occurs (it is independent on time).
Besides, it is assumed that the emission probabilities are also homogeneous.

\item Conditional independence among the observations: each observation $x_i$ only depends on the state $y_i$, that is $x_i$ is generated by the state $y_i$ with $\mathbb{P}\left(x_i \vert y_i \right)$.
%
\end{itemize}

\subsection{Applying HMM in Biological Contexts}
%
%
%
A DNA is a macromolecule composed of simple units called nucleotides. 
Each nucleotide is composed of deoxyribose, phosphate and one of the four nitrogenous bases: \textit{adenine (A)}, \textit{cytosine (C)}, \textit{guanine (G)}, \textit{thymine (T)}.
The DNA is represented by a double helical structure wherein each complementary base pair (\textit{A-T, C-G}) are connected by hydrogen bonds.
The two DNA strands are in opposite directions to each other, positive or $5'-3'$ (left to right) is complemented by the reverse or complementary sequence.
The selection of a HMM is given by the alphabet, the number of states and the pattern of transitions of the Markov chain.
For this problem, the alphabet is defined as ${\cal{A}}=\{A,C,G,T\}$. The number of states depends on the problem, for example can be 
${\cal E}=\{\text{gene}, \text{intergenic region}\}$ if the goal is determining 
genes and intergenic regiones, and can be ${\cal E}=\{\text{exon}, \text{intron}, \text{intergenic region}\}$ if also there is interest in identifying non-coding (introns) regions. Although, the hidden space can be easily generalised for more states in the case that we are interested in identifying a higher number of regions in the sequence. The emission and transitions probabilities are numerically computed.
%

%
%
%
%
%

%
More formal, given an alphabet ${\cal A}$, a set of states ${\cal E}$ and a DNA sequence  $x$ composed by $x_1, x_2, \ldots, x_n$ where $x_j\in{\cal{A}}$, the task is to assign to each symbol $x_j$ for $j=1,\ldots,n$ one label $y_i$ in ${\cal E}$.
In the specific problem of VSG-gene identification, we define the alphabet as ${\cal{A}}=\{A,C,G,T\}$, and we analyse the HMM model for two situations: considering two states (${\cal E}=\{\text{VSG-gene}, \text{no VSG-gene}\}$) and considering three states\\ 
(${\cal E}=\{\text{VSG-gene}, \text{no VSG-gene}, \text{other structures}\}$). The sequence $x$ was taken from the public \textit{GenBank} database~\citep{Benson09}.

%
%
%
%

%

%
%


\subsection{Related Works}
HMMs have been applied in several areas of pattern recognition due to their powerful for modelling sequential data.
In particular, they have been especially used for developing speech recognition devices~\citep{Rabiner89}.
Basically, this problem consists in predicting a spoken word from a speech signal and translate into text.
For example, hybrid techniques that use ideas from the Neural Network domain and HMMs has been applied for speech recognition systems~\citep{Trentin01}.
Another tool used for solving this problem has been Gaussian Mixture HMM~\citep{Rabiner89}.
In the case of phone speech recognition a hybrid model that employs deep learning with HMMs was studied in~\citep{Dahl11}.
HMM has been also applied for acoustic modeling using Mixture of Gaussian Distributions~\citep{Juang86,Rabiner05}.
In addition, the model was successfully applied in the Human-Computer Interface area. 
In~\citep{Yamato92} time sequential images expressing human actions were encoded in a sequence of symbols, which were modeled by a HMM. 
In~\citep{Liu11} the authors use the sequence of facial temperature for emotion recognition, the transitions between emotional states are modelled by HMM.

Since the late 80s, HMM has been successfully applied in the area of computational biology~\citep{Durbin98}. 
In~\citep{Churchill89} the author evaluated HMM for modeling five DNA sequences:
the yeast mitochondrial, human and mouse mitochondrial, a human X chromosomal fragment, and the bacteriophage lambda genome.
Other applications in this first stage have been presented in~\citep{Krogh94} and~\citep{Baldi94}. In~\citep{Krogh94} the authors applied HMM for searching common patterns between a protein sequence and multiply aligned sequences.
They shown the effectiveness of the model for searching a sequence in a given protein family.
In~\citep{Baldi94} the HMM was used to capture statistical characteristics in three protein families: globins, immunoglobulins, and kinases, wherein the model was applied for classification, multiple alignments, and motif detection.
Another protein structure modeling using HMM was presented in~\citep{Stultz93}.
A classification problem was studied in~\citep{Henderson96} where the authors applied HMM for segmenting uncharacterised genomic DNA sequences into exons, introns and intergenic regions.
The DNA sequencing errors were modeling in~\citep{Lottaz03} where the authors improved the detection of translation start and stop sites, and thus the location of the coding regions.

Other applications correspond to the task of searching structures on genomes. The model was used for gene searching in bacterial genomes~\citep{Lukashin98}, and in~\citep{Delcher99} was shown the effectiveness of HMM for finding gene on eukaryotic genomes. The model was also used for identifying homologous protein and nucleotide sequences~\citep{Finn11}.
A variation of HMM named~\textit{Profile-HMM} have been applied for analysing genomic sequences~\citep{Eddy95, Eddy96, Eddy98, Gough01}.
The huge interest in the community concerning HMMs for solving biological computational problems has caused that several researchers develop specific software for gene sequence and prediction signals, these include~\citep{Lukashin98,Delcher99, Finn11,Flicek07}.

Other learning techniques have also been used for solving classification tasks and clustering problems in the field of computational biology. 
In~\citep{DeCaprio07} the authors described a gene predictor based on \textit{Conditional Random Fields (CRF)} that incorporates information of \textit{Expressed Sequence Tags (ESTs)} for predicting genes on fungal genome. 
In~\citep{Gross07} the authors combined CRF with \textit{Support Vector Machines (SVM)} and tested for the human genome, the model incorporates data from other genomes called informants, making a multiple alignment based on some assumptions about the respective evolutionary processes.
Another application of the HMM combined with SVM analyses the genome of a nematode~\citep{mgene09}.

In biological sequence analysis \textit{Artificial Neural Network (ANN)} have been also used. In~\citep{Rebello11} ANN was applied for predicting genes on a Streptococcus genome, and they were also used on the prediction of splice sites of a gene in~\citep{raey09}. \textit{Decision trees} were performed for finding genes in vertebrate DNA sequences~\citep{Salzberg96}, and a variation (called \textit{Randomized oblique decision trees}) combined with other learning techniques were applied for gene modeling in~\citep{Allen04}.


In the particular case of  \textit{T. brucei}, an evolutionary analysis considering the family members of this gene was introduced in~\citep{Alvarez96}. Although, in this article the authors analyse the sequence doing pairwise comparisons between \textit{T. brucei} and other protein coding genes. An example of using a system based on HMM to search genes on a particular \textit{T. brucei} chromosoma is presented in~\citep{El-sayed03}, although in this article the sequence has only one VSG cluster.
%

During the last decades, HMM have been used in a large number of applications for analysing sequential data. Here, we presented a summary that is not exhaustive, a more complete overview about HMM applications in biological sequences can be seen in~\citep{Durbin98,Choo04, Yoon09}.
%
%

%


%
\section{Training and Inference}
\label{Training}
%
Let $\bm{\theta}$ be the vector with the parameters of the model (the probabilities of transitions and the probabilities of emissions).
The training process is based in the \textit{Maximum-likelihood Estimation (MLE)} that consists in adjusting parameters to maximize the likelihood for a given sequence $x$ generated by the model. That is finding $\bm{\theta}$ such that $\mathbb{P}_{\theta}(x)$ is maximised. 
For any two states $k$ and $l$, and a symbol $b$ we find the parameters $p_{kl}=\mathbb{P}\big(y_{i+1}=l| y_i=k\big)$ and $e_{k}(b)=\mathbb{P}\big(x_{i}=b| y_i=k\big)$ as follows:
$$\hat{p}_{kl}=\frac{P_{kl}}{\sum_{l'}P_{kl'}}, \qquad \hat e_k(b)=\frac{E_k(b)}{\sum_{b'} E_k(b')},$$
where $P_{kl}$ is the number of transitions from $k$ to $l$, and $E_k(b)$ is the number of times that the state $k$ emits the symbol $b$ in the current training sequence.
The adjustable vector parameter $\theta$ is a collection composed by $\hat{p}_{kl}$ and $\hat{e}_k$ for all $k$ and $l$.


A computational efficient algorithm for setting the parameter is \textit{Baum-Welch (BM)} Algorithm~\citep{Welch03Lecture}, that is a particular case of the well-known \textit{Expectation-Maximization (EM)}~\citep{Baum70, Dempster77,LibroHastie}.
The BW algorithm is based in two auxiliary recursive procedures called \textit{forward} and \textit{backward} algorithms. 
%
%
The forward algorithm is a dynamic programming algorithm that computes in an efficient way the \textit{forward probability}, which is given by the following expression~\citep{Yoon09}
\begin{eqnarray*}
\alpha_k(n)&=&\mathbb{P}(x_1,\ldots,x_n,y_n=k)\\
&=&\mathbb{P}\big(x_{n}| y_n=k\big)\sum_{j}\alpha_j(n-1)\mathbb{P}\big(y_{n}=k| y_{n-1}=j\big).
\end{eqnarray*}
The time-complexity of the forward algorithm is $nK^{2}$ where $K$ is the number of states and $n$ is the size of the sequence $x$.
Algorithm~\ref{ForwardAlgo} presents the procedure used for compute the forward probabilities $\alpha_k(n)$ for all states $k$. 

\begin{algorithm}[h!]
\caption{\label{ForwardAlgo} Forward Algorithm.}
\tcp{Initialization}
	$\alpha_k(1)=\mathbb{P}\big(x_1, y_1=k\big)=\mathbb{P}\big(y_1=k\big)	\mathbb{P}	\big(x_1\vert y_1=k\big)$, $\forall k=1,\ldots K$\;
	\For{($i=1$ to $n$)}{
		$\displaystyle{\alpha_{k}(i)=\mathbb{P}\big(x_{i}| y_i=k\big)\sum_{j}\alpha_j(i-1)\mathbb{P}		\big(y_{i}=k| y_{i-1}=j\big)}$\;
	}
	$\displaystyle{\mathbb{P}(x)=\sum_{k}\alpha_k(n)}$\;
	Return $\mathbb{P}(x)$ and $\alpha_k(n)$ for all $k$\;
\end{algorithm}
%
%
The backward probability is also recursively defined
\begin{eqnarray*}
\beta_k(n)&=& \mathbb{P}(x_{n+1},\ldots,x_n|y_n=k)\\
&=&\sum_{l}\mathbb{P}\big(x_{n+1}| y_{n+1}=l\big)\beta_l(n+1)\mathbb{P}\big(y_{n+1}=l| y_n=k\big).
\end{eqnarray*}
Algorithm~\ref{BackwardAlgo} illustrates the procedure used for compute the backward probabilities $\beta_k(n)$ for all states $k$. 
In a similar way that the forward case, this backward probability can be recursively computed in a $nK^2$ time-complexity~\citep{Yoon09}.
\begin{algorithm}[h!]
\caption{\label{BackwardAlgo} Backward Algorithm.}
	\tcp{Initialization}
	$\beta_k(n)=1$  $\forall k=1,\ldots K$\;
	\For{($i=n-1$ to $1$)}{
		$\displaystyle{\beta_{k}(i)=\sum_{l}\mathbb{P}\big(x_{i+1}| y_{i+1}=l\big)\beta_l(i+1)\mathbb{P}\big(y_{i+1}=l| y_i=k\big)}$\;
	}
$\displaystyle{\mathbb{P}(x)=\sum_{k}\mathbb{P}(y_1 \vert y_0=k)\mathbb{P}\big(x_{1}| y_1=k\big)\,\beta_k(1)}$\;
	Return $\mathbb{P}(x)$ and $\beta_k(n)$ for all $k$\;
\end{algorithm}
%


\subsection{Description of the Baum-Welch Algorithm}
\label{BW}
The BW algorithm is iterative, at the first iteration is initialised the transition probability matrix $\trans$ and the emission matrix $\emis$. 
The initialisation can be done using a uniform distribution or in an arbitrary way based on \textit{a priori} known information.
In the following we present in brief the main mathematical expressions of the Baum-Welch method.
%
%
%
Let $x$ be the current training sequence, the transition probability from the state $k$ to the state $l$ at the position $i$ of $x$ is computed as~\citep{Yoon09}
$$\mathbb{P}\left(y_i=k,y_{i+1}=l\vert x\right)=\frac{\alpha_k(i)\mathbb{P}\big(y_{i+1}=l| y_i=k\big)\mathbb{P}\big(x_{i+1}| y_{i+1}=l\big)\,\beta_l(i+1)}{\mathbb{P}\left(x\right)},$$
where ${\mathbb{P}\left(x\right)}=\displaystyle{\sum_{k}{\alpha_k(n)}}$ is computed using the forward algorithm~\citep{Rabiner89}.
%
%
The probability of generating the symbol $b$ when the state is $k$ is computed as 
\begin{equation}
\displaystyle{\frac{\sum_{\{i:x_{i}=b\}}\alpha_{k}(i)\beta_{k}(i)}{\mathbb{P}(x)}}.
\end{equation}
Let $T$  be the number of sequences in the training set, and we denote the training sequences by $x^{(t)}$ for $t=1,\ldots,T$.
The probabilities of transition and emission are computed as
%
%
\begin{equation*}
\displaystyle{\hat{p}_{kl}=\frac{P_{kl}}{\sum_{l'}P_{kl'}}}
\qquad
\rm{and}
\qquad
\displaystyle{\hat{e}_k(b)=\frac{E_k(b)}{\sum_{b'} E_k(b')}},
\end{equation*}
where
\begin{equation}
P_{kl}=\sum_{t}\frac{1}{\mathbb{P}(x^{(t)})}\sum_{i}\alpha^t_k(i)\mathbb{P}\big(y_{i+1}=l| y_i=k\big)\mathbb{P}\big(x^{(t)}_{i+1}| y_{i+1}=l\big)\,\beta^{(t)}_l(i+1),
\end{equation}
\begin{equation}
E_k(b)=\sum_{t}\frac{1}{\mathbb{P}(x^{(t)})}\sum_{\{i:x^{(s)}_{i}=b\}}\alpha^{(t)}_{k}(i)\beta^{(t)}_{k}(i).
\end{equation}
It has been proven in~\citep{LibroHastie} that the likelihood function increases at each step of the BW algorithm. That is $\displaystyle{P_{\hat \theta^{(h+1)}}(x)\geq P_{\hat \theta^{(h)}}(x)}$ for all step $h$.
The algorithm ends when the difference between the likelihood function at the step $h+1$ and the value at the step $h$ is lower than some arbitrary threshold, that is $\displaystyle{dist(P_{\hat \theta^{(h+1)}}(x), P_{\hat \theta^{(h)}}(x))<\epsilon}$ where $dist(\cdot)$ is an arbitrary distance. In our numerical experiences, we used Euclidean distance and $\epsilon=0.00001$, in addition we control the number of iterations (the maximum number of iterations was 500). 

\subsection{Description of the Viterbi Algorithm}
\label{Viterbi}

Once the model parameters are estimated, the Viterbi algorithm is useful for finding the \textit{best} sequence of states $y$ for generating a specific sequence $x$.
The method assumes that the \textit{best} sequence is that that generates the sequence $x$ with higher probability, that is
\begin{equation}
\label{bestPath}
y^{*}=\arg\max_{y}\frac{\mathbb{P}(x,y)}{\mathbb{P}(x)}.
\end{equation}
Note that the denominator in~(\ref{bestPath}) does not depend of $y$. So, the goal is to find $y^{*}$ such that  $\mathbb{P}(x,y)$ is maximised.
Viterbi algorithm is iterative. Given the samples until the position $i$, let $V_{i}(k)$ be the probability of the best sequence of states (\textit{path}) that generates the state $k$.
At the initial iteration, the method sets  $V_1(k)$ with the probability that observation $x_1$ is generated.
%
Next, $V_{i}(l)$ is computed for each position $i=2,\ldots,n$ and each state $k=1,\ldots,K$.
At the step $i$, the algorithm finds the sequence of states $y_1, y_2, \ldots, y_i$ that generates the sequence $x_1,x_2, \ldots ,x_i$ with highest probability,  for each location $i$ and state $k$ ($y_i=k$). 
The algorithm computes the $V_{i+1}(l)$ for the location $i+1$ and all possible states $l$ following 
\begin{equation}
\label{Vi}
V_{i+1}(l)=\max_{k}\bigg(V_i(k)\,\mathbb{P}\big(y_{i+1}=l| y_i=k\big)\bigg)\mathbb{P}\big(x_{i+1}| y_{i+1}=l\big).
\end{equation}
We define an auxiliary variable $I_i(l)$ that contains the state that maximises the expression~(\ref{Vi}) at each iteration.
The probability $\mathbb{P}(x,y^{*})$ is given by the maximum value of $V_n(k)$ for all the states $k$. 
The most probable path $y^*$ is computed using a {backtracking procedure} from ($n$ till $1$) and $y_{i-1}^*=I_i(y^*_{i})$.
%


%

\subsection{Technical Issues}
We present here some technical issues related to the performance and numeric stability of the algorithms during the training process
\begin{itemize}
\item Algorithmic stability:
The product among probabilities should be controlled, because it can occur underflow.
This is a common problem when we are working with power of Markov matrices and products of probabilities. The stability can be improved using scaling coefficients in the forward and backward probabilities and logarithmic transformations in the Viterbi algorithm~\citep{Durbin98, Rabiner89}. 
\item Performance of learning on imbalanced dataset: we say that a dataset is imbalanced if the classes are not \textit{approximately equally represented}~\citep{Chawla02}. This imbalance can provoke suboptimal classification performance, due to the learning tool overweight the large classes ignoring the small ones~\citep{Chawla02}.
Some techniques have been introduced for improving the predictors in the imbalanced context, which can be performed for solving the gene classification problem~\citep{Chawla04}.
%

%

%
\end{itemize}

%

\section{Methodology}
\label{Methodology}
%
In this section we describe the methodology used for applying HMM for solving the problem of the labelling areas of  \textit{T. brucei} genome as VSG gene or non-VSG gene.
%
The alphabet is composed by the four nucleotide types ${\cal{A}}=\{A,C,G,T\}$.
Concerning the number of states in the problem, we begin by adjusting the model with two states, which represents a binary situation: yes or not \textit{VSG} region.
Let $S$ be the non-redundant collection of sequences of \textit{T. brucei} genome. 
The set $S$ has~$9$ sequences of different lengths and different numbers of clusters of VSG genes.
%
%
As usual, we generate a random partition of the set $S$ in two subsets: a training set $S^{T}$ and a validation set $S^{V}$.
The sequences in $S^{T}$ are used for estimating the transition probability matrices and the parameters of the emission distribution.
We apply BW Algorithm for performing that estimation.
Note that, only the nucleotide sequence is taken account in the BW Algorithm, the information about the VSG genes is not used.
Next, we apply the Viterbi Algorithm on the sequences in $S^{V}$ in order to estimate the most likely sequence of hidden states. 
We determine those regions in the testing sequences that are predicted as VSG genes. 
Finally we evaluate the  model using the predicted values with the VSG annotated genes of the sequences in $S^{V}$.

\begin{figure}[h]
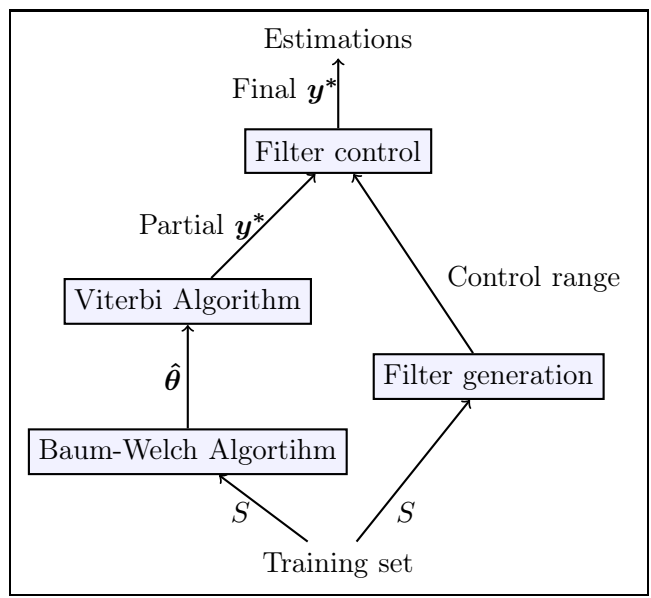

\begin{center}
\fbox{
	\tikz{
		\node[fill=white] (ts) at (0,-0.5) {Training set};
		\node[draw,thick,fill=blue!5, rectangle,minimum size=3.5ex]  (ba) at (-2,1) {Baum-Welch Algortihm};
		\node[draw,thick,fill=blue!5, rectangle,minimum size=3.5ex]  (va) at (-2,3) {Viterbi Algorithm};
		\node[draw,thick,fill=blue!5, rectangle,minimum size=3.5ex]   (f) at (2,2) {Filter generation};
		\node[fill=white] (e) at (0,6.5) {Estimations};
		\node[draw,thick,fill=blue!5, rectangle,minimum size=3.5ex]  (fc) at (0,5) {Filter control};
		\node[fill=white] (l1) at (-1.3,0.2) {$S$};
		\node[fill=white] (l2) at (0.9,0.2) {$S$};
		\node[fill=white] (l3) at (-2.2,2) {$\bm{\hat{\theta}}$};
		\node[fill=white] (l3) at (-1.8,4) {Partial $\bm{y^*}$};
		\node[fill=white] (l3) at (-0.7,5.8) {Final $\bm{y^*}$};
		\node[fill=white] (l3) at (2.6,3.3) {Control range};
		\draw[->,black] (ts) edge[->,thick] (ba);
		\draw[->,black] (ba) edge[->,thick] (va);
		\draw[->,black] (ts) edge[->,thick] (f);
		\draw[->,black] (f) edge[->,thick] (fc);
		\draw[->,black] (va) edge[->,thick] (fc);
		\draw[->,black] (fc) edge[->,thick] (e);			
		}
	}
\caption{\label{SchemaProcedure}Schema of the procedure for gene sequence classification using HMM. A training set composed by sequences is the input of the Baum-Welch algorithm, then the Viterbi algorithm is applied producing a \textit{partial} estimation. The training sequence is also used for generating an interval control corresponding the locations of the VSG genes. Next, the location interval of control is applied as filter on the \textit{partial} estimation to produce the final estimation.}
\end{center}
\end{figure}

In our experiments we find that the predictor can be improved defining an interval of control, which we call \textit{location control filter}.
%
We can define a range where the VSG are presented, which is based on the minimum and maximum lengths of the labeled VSG.
Once this range is defined, we can use it as filter for the predictions given by the BW and Viterbi algorithms.
The model assigns a label VSG to a sub-sequence of nucleotide if verifies the following two conditions: it is predicted by the HMM model such as VSG and  it passes the filtering operation.
We call \textit{filter operation} if the subsequence has location inside the range defined by the training set.
As a consequence, it is possible to adjust the predictions given by the model discarding those that do not pass the filter. 
Obviously, the range depends of the training set, then it can change according different training data.
In spite of that, this is a common practice in machine learning. For instance, when the input patterns are normalised. This normalisation more often takes as reference the maximum and minimum value of the input variables. 
Here, the approach is similar, we consider the minimum VSG location and the maximum VSG location using the training data, out of this range the genes are considered non-VSG.
Figure~\ref{SchemaProcedure} presents a hierarchical schema among the techniques that compose our approach.


We evaluate the learning capability of the model using the following quantitative metrics.
Table~\ref{ConfusionMatrix} shows the error types (confusion matrix) that can happen during the parameter  estimations, four situations are identified:
\begin{itemize}
\item  \textit{True positive (TP)} refers when the VSG was correctly classified.
\item \textit{False positive (FP)} refers to an incorrectly  identification, in our problem that occurs when the HMM predicts that some nucleotides are part of the VSG gene when this is not correct.
\item \textit{False negative (FN)} refers when the model incorrectly rejected a VSG gene.
\item \textit{True negative (TN)} refers when the model correctly rejected a VSG gene.
\end{itemize}
We measure the quality of our estimation using two well-known measures of binary prediction quality: the 
\textit{True Positive Rate (TPR)} (also named \textit{sensitivity}) and the \textit{Positive Predictive Value (PPV)} (also named \textit{precision}).
The TPR is the proportion of nucleotides correctly  predicted  over the total number of nucleotides that are part of VSG gene. It is defined as:
\begin{equation}
TPR=\displaystyle{\frac{TP}{TP+FN}}.
\end{equation}
The PPV measures the proportion of correctly classify nucleotides over the total of VSG identifies. It is given by:
\begin{equation}
PPV=\displaystyle{\frac{TP}{TP+FP}}.
\end{equation}

\begin{table}[h!t]
\centering
\caption{Confusion matrix: type of estimation errores.\label{ConfusionMatrix}}
\begin{tabular}{|c||c|c|}
\hline
Prediction $\diagup$ Target & Positive & Negative  \\
 \hline\hline
Positive & True Positive (TP) & False Positive (FP)  \\
 \hline
Negative & False Negative (FN)  & True Negative (TN)  \\
 \hline
\end{tabular}
\end{table}

%

The following hypothesis are considered. We assume that exists independence among sequences, as well as among genes into the sequences. That is, we dispose of a non-redundant collection of sequences of \textit{T. brucei}  genomic data.
%
%
A common problem when we are working on pattern recognition for genome sequences is that we can not affirm the independence between genes and segment of genes in the sequence.
Some parts in the sequence can be strongly correlated each of other.
The correlation among several sequences arises from the process of phylogenetic inertia, basically this means that the sequences have the same origin.
In spite of that, the process of divergence evolution can vanish the phylogenetic information. Even two sequences that diverged from a common ancestral sequence can absolutely lost the information from their ancestors if the sequences diverged enough.
Often, it is used the percentage of identical base pair (sequence identity) between two sequences as correlation assessment between that sequences. 
As a consequence, in order to assume independence among sequences and to produce a non-redundant collection, the sequence percent identity is used as independence assessment reference.
The VSG genes diverge from a common ancestor, despite that their amino acid identity is low. An usual assumption is considering the VSG genes as non-redundant set, and assuming the  hypothesis of independence among the sequences.



\section{Results}
\label{Result}
\subsection{Data description}

We use a set of 9 genomic sequence segments from chromosome 9 of  \textit{T. brucei} genome where some VSG gene clusters are located. 
The data set was taken from the \textit{GenBank} database~\citep{Benson09}.
%
Table 2 shows size of sequences and number of annotated VSG genes for each one.
Since in these genomic sequences the location of most VSG had been previously identified using traditional homology-based annotation methods, our predictions obtained with the HMM approach  can be  compared with these annotations obtained by traditional based methods.
\begin{table}[h!t]
\centering
\caption{Size of the sequence segments and number of annotated VSG genes in each sequence of chromosome 9 of  {\it T. brucei}.\label{secuencias_cr9}}
\begin{tabular}{c c c}
\hline\hline
Sequence Id & Size & Number of VSG genes\\
 \hline
$S_1$ & 75462  & 16 \\
$S_2$ & 67530  & 21 \\
$S_3$ & 19745  & 7  \\
$S_4$ & 89317  & 10 \\
$S_5$ & 102962 & 3 \\
$S_6$ & 17110  & 5 \\
$S_7$ & 88923  & 8  \\
$S_8$ & 24879  & 0  \\
$S_9$ & 65933  & 7  \\
\hline\hline
\end{tabular}
\end{table}
\subsection{Applying the HMM model with $2$ states}
We start by analysing the performance of the HMM model with $2$ states.
Without loss of generality we index the sequences in $S$ from $1$ to $9$ as $S_1,S_2,\ldots,S_9$. 
We create three partitions of $S$ using a uniform random selection. Each partition have two groups, one is used for training and another one for validations.
Table~\ref{TrainVal} shows the training and validation sets for each partition of $S$.
In case of HMM with 2 states the BW  algorithm converges after $160$ iterations for the three training sets.
We compute the Viterbi path and the model accuracy.
\begin{table}[h!t]
\centering
\caption{Generation of the training and validation sets.}\label{TrainVal}
\begin{tabular}{c c c}
\hline\hline
Partition Id & Training set & Validation Set\\
 \hline
$A$ & $\{S_1, S_2, S_4, S_5, S_7, S_8\}$  & $\{S_3,S_6,S_9\}$\\
$B$ & $\{S_1, S_2, S_3, S_4, S_6, S_9\}$  & $\{S_5,S_7,S_8\}$\\
$C$ & $\{S_3,S_5,S_6,S_7,S_8,S_9\}$  & $\{S_1,S_2,S_4\}$\\
\hline\hline
\end{tabular}
\end{table}

We reached the following estimated matrices when we used the sequences of the partition $A$
\begin{center}
$\hat{\trans}=\left(\begin{array}{cc}
         0.9969  &  0.0031\\	
         0.0029  & 0.9971\\	
\end{array}
\right),$ 
\end{center}
where the $(i,j)$ element represents the transition probability from state $i$ to $j$ for $i,j=1,2$ and 
%
 \begin{center}
$\hat{\emis}=\left(\begin{array}{cccc}
0.3815  &  0.1911 &  0.2018 &  0.2255\\
0.1932  &  0.2213 &  0.2031 &  0.3824\\		
 \end{array}
\right)$
\end{center}
where $(i,1)$, $i=1,2$ represents the probability of state $i$ emits the symbol $A$, $(i,2)$ the probability that  state $i$ emits the second symbol $C$ and so on.
The matrix of transition probabilities is diagonal dominant, therefore the chance that the model jumps from one states to another one is very low.
Table~\ref{SP_SN_1} shows some metrics of the accuracy of our model.
The rows of the table show the accuracy metric reached for the model with 2 states.
The first column specify the metric, the next three columns shows the TPR, PPV, and  number of VSG non-detected (ND-VSG) using the partition $A$. 
The other following columns present the results using $B$ and $C$ sets.
The model predicts all VSG genes in almost all cases, therefore TPR is high.
In the worst case the TPR is higher than $0.81$. 
The PPV for the sequences $S_9$ of the partition $A$ and $S_5$ of the partition $B$ are small, as a consequence the model prediction has a relatively high false positive error.
\begin{table}[h!]
\centering
\caption{Sensitivity (TPR), precision (PPV) and number of non-detected VSG (ND-VSG) reached by a HMM with 2 states}.\label{SP_SN_1}
\begin{tabular}{ l c c c c c c c c c}
  \hline\hline
$\;$  & $\;$ & $A$ & $\;$ & $\;$ & $B$ & $\;$ & $\;$ & $C$ & $\;$\\  
$\;$  & $S_3$ & $S_6$ & $S_9$ & $S_5$ & $S_7$ & $S_8$ &$S_1$ & $S_2$ & $S_4$\\
\hline
TPR  & 0.9709  & 0.9634 & 0.9838 & 0.9581  & 0.9485  &  - & 0.9759  &  0.8104 & 0.9959\\
PPV  & 0.5214 &  0.4652 & 0.2781  & 0.1769  & 0.4775  &  -& 0.3171  &  0.4455 & 0.3085\\
ND-VSG  & 0  &  0 & 0 & 0  &  0  & 0 & 8  &  4  & 0\\
\hline\hline
\end{tabular}
\end{table}

\subsection{Application of the HMM with $3$ states}
In order to improve the accuracy in predictions, we analyse a HMM with 3 states. Our purpose is to investigate the existence of another group that does not appear in the HMM with 2 states.
We use training and validation set presented in Table~\ref{TrainVal}.
The BW algorithm converges for the three partitions in less than $400$ iterations.
%
%
%
  
\begin{figure}[h!t]
\begin{center}
\includegraphics[scale=0.5]{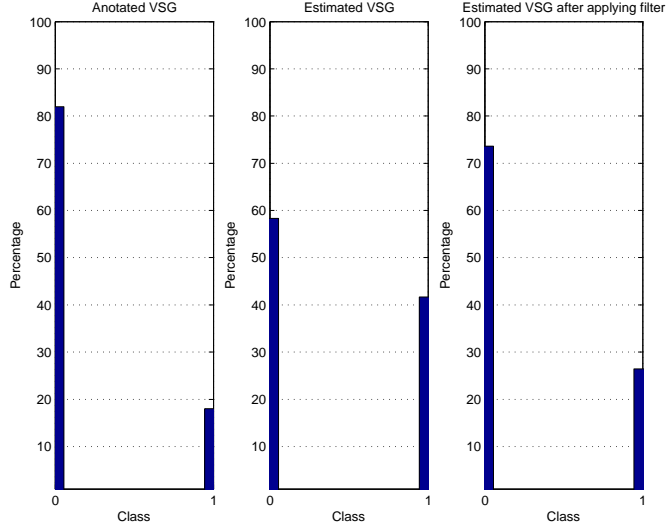}
\end{center}
\caption{\label{Histograms}
Example of the improvement for controlling the gene location. The data corresponds to the sequence 9. Left histogram was generated with the original data, in the center the histogram was done with the estimations produced by the HMM, and the right histogram corresponds to the estimations with a location control.}
\end{figure}

%
The estimated matrix of transitions when was used the sequences of the partition $A$ was,
\begin{center}
$\hat{\trans}=\left(\begin{array}{ccc}
         0.9956 & 0.0028 & 0.0016\\	
         0.0021 & 0.9976 & 0.0003\\	
         0.0017 & 0.0003 & 0.9980\\
\end{array}
\right),$ 
\end{center}
where the $(i,j)$ element represents the transition probability from state $i$ to $j$ for $i,j=1,2,3$.
As well as in the case of a model with two states, that matrix is diagonal dominant.
The estimated matrix of emissions when was used the sequences of the partition $A$ was,
 \begin{center}
$\hat{\emis}=\left(\begin{array}{cccc}
0.3610 & 0.1455 & 0.1388 & 0.3547\\		
0.1712 & 0.2346 & 0.2267 & 0.3675\\		
0.3610 & 0.2330 & 0.2369 & 0.1691\\		
 \end{array}
\right),$
\end{center}
where the first column of the matrix corresponds to the base $A$, the second to the base $C$, the third one to the base $G$ and the last one to the base $T$, so the ($i,1)$ represents the probability of state $i$ emits the symbol $A$ and so on, $i=1,2,3$.
Table~\ref{SP_SN_2} presents the accuracy of the model with 3 states.
The first three columns shows TPR, PPV and number of non-detected VSG genes for partition $A$. The next group of three columns present accuracy for partition $B$ and the last columns corresponds to partition $C$.
We can affirm that the model with 3 states fits better than the 2-state model.
The TPR metric is slightly lower in this case, but the precision increases for all sequences.
\begin{table}[h!t]
\centering
\caption{Sensitivity (TPR), precision (PPV) and number of non-detected VSG (ND-VSG) reached by a HMM with 3 states.}
\begin{tabular}{l c c c  c c c  c c c}
  \hline\hline
$\;$  & $\;$ & $A$ & $\;$ & $\;$ & $B$ & $\;$ & $\;$ & $C$ & $\;$\\  
$\;$  & $S_3$ & $S_6$ & $S_9$ & $S_5$ & $S_7$ & $S_8$ & $S_1$ & $S_2$ & $S_4$\\\hline
TPR      & 0.9472  & 0.9540 & 0.9559 & 0.9535  &  0.9405 & - & 0.9559 & 0.9280 & 0.9403\\
PPV      & 0.6712 & 0.7080  & 0.4022 &  0.3985 &  0.6629 & - & 0.5027 & 0.5435 & 0.4557\\
ND-VSG &      0  &     0 &       0 &       0 &      0  & 0  & 0 & 0 & 0 \\
\hline\hline
\end{tabular}
\label{SP_SN_2}
\end{table}
Table~\ref{pred_seq9} shows the locations of annotated VSG (left column) and predicted VSG (right column) by the HMM model with 3 states using the partition $A$ and sequence $9$. 
The model correctly predicts seven VSG genes and misclassifies other seven ones.
%
%
The length of VSG genes correctly predicted in the sequence $9$ varies from a minimum of $1115$ up to $2217$.
The VSG genes wrong predicted by the model have lengths: $113$, $510$, $653$, $250$, $206$, $5600$ and $2432$. Note that all the bad predictions are out of the range $[1115,2217]$. 
This situation is repeated in a large number of wrong predicted genes on the validation sequences.
As a consequence, if we consider the lengths of the annotated  VSG genes (in the training set) we can define a bounded range by the minimum and maximum length of well-predicted VSG genes.
Then, it is possible to adjust the predictions provided by the model discarding those that are outside of this range.
Therefore, we can use the training set for both: to set the parameters of the model and to define a range for the VSG length.
Figure~\ref{Histograms} illustrates how the model is improved when a location filter is performed.
Besides, it shows an additional challenge of the learning model due to the imbalance among the classes~\citep{Chawla04}.
There are three histograms, the horizontal axe indicates the classes and the vertical axe indicates the percentage of genes. The data is the testing and corresponds to the sequence 9. There are two classes: the non-annotated VSG (denoted by $0$) and the annotated VSG (denoted by $1$). 
The left histogram shows the testing data, in the center is presented the estimations produced by the HMM without control of the gene locations, and the right histogram presents the estimations produced by the HMM with location control.
The serie of histograms shows how is improved the percentage of well-classified genes for the sequence 9 using the location filter, the increment is larger than $10\%$ of cases.

\begin{figure}[h!t]
\begin{center}
\includegraphics[scale=0.6]{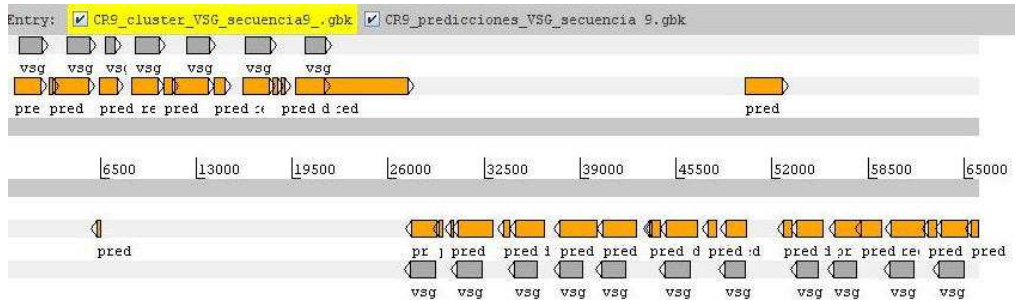}
\end{center}
\caption{\label{artemis} Annotated VSG genes and their predictions by an HMM with 3 states for the sequence 9. The Figure was generated using Artemis software~\citep{Artemis2}.}
\end{figure}

The VSG genes in the sequence~$9$ are located between the first~$22000$ nucleotides (Table~\ref{pred_seq9}). 
In  this region, the most likely path of states determined by the Viterbi algorithm shows that only states $1$ and $3$ are presented (the state $3$ corresponds to the VSG genes), with only the exception between the coordinates~$6277-6429$.
From~$27560$ until the end of the sequence, the changes are between states $1$ and $2$ (with only the exception between $50143$ and $52575$ locations). 
This indicates that in addition to the VSG genes, the model detects other regions (corresponding to state $2$) with particular characteristics that make them distinguishable from those predicted as states $1$ and $3$.
We observe that those particular regions located at the second middle of the sequence and relating to the state $2$ correspond to the VSG genes in the complementary sequence (Table~\ref{pred_seq9_compl}).
Table~\ref{pred_seq9_compl} presents an example about the location of annotated (left column) and predicted (right column) VSG genes by the 3-states HMM for the complementary sequence 9.
Figure~\ref{artemis} was generated with the Artemis Software~\citep{Artemis,Artemis2}. It shows the location of annotated VSG genes and predicted by the 3-state model for sequence 9.
The annotated VSG genes are indicated with the label ``vsg'' and the predictions are labeled as ``pred''.
The figure is useful for explain the model accuracy in finding the VSG on the positive strand (left to right direction, located at the top of the figure) and complementary strand (opposite direction, situated on the bottom) for the sequence 9.
Table~\ref{Complementary} shows the performance reached by the HMM with 3 states for the complementary sequence. The table presents the TPR and PPV values and the number of not detected VSG. The three groups of columns correspond to the $A$, $B$ and $C$ learning partitions.
We can see that all VSG genes located on the complementary sequence are well-identified by the model.
\begin{table}[h!t]
\centering
\caption{Location of the VSG genes annotates and predicted by the model with 3 states. The data corresponds to the sequence 9.\label{pred_seq9}}
\begin{tabular}{c c}
\hline\hline
Location of the VSG gene & Location of the VSG prediction\\
\hline
1001 - 2502  & 658 - 2427\\
$\;$&3056 - 3169\\
4223 - 5692  &3423 - 5640\\
6846 - 7375  &6429 - 7544\\
8875 - 10387 &8626 - 10315 \\
             &10829 - 11339\\
12360 - 13819&11555 - 13748\\
             &14192 - 14845\\
16342 - 17943 &16184 - 17849\\
             &18148 - 18398\\
             &18712 - 18918\\
20343 - 21634 &19746 - 21585\\
             &21676 - 27276\\
             &50143 - 52575\\
\hline\hline
\end{tabular}\end{table}
\begin{table}[h!t]
\centering
\caption{
Location of the VSG gene and their predictions for the HMM with 3 states for the complementary sequence. The data corresponds to the sequence 9.}\label{pred_seq9_compl}
\begin{tabular}{c c}
\hline\hline
Location of the VSG gene & Location of the VSG prediction\\
\hline
27500 -	29086 & 27560 -	29217\\
              & 29404 -	29594\\
              & 30216 -	30348\\
30672 -	32290 & 30752 -	32972\\
              & 33791 -	34145\\
34509 -	35973 & 34619 -	36483\\
37592 -	39046 & 37648 -	40083\\
40462 -	41986 & 40528 -	42688\\
              & 43715 -	43783\\
              & 43853 -	44335\\
44756 -	46342 & 44816 -	46881\\
              & 47692 -	48183\\
48819 -	50113 & 48910 -	50143\\
              & 52782 - 53268\\
53634 -	55059 & 53685 -	55339\\
56172 -	57665 & 56223 -	57969\\
              & 58034 -	59358\\
59975 -	61544 & 60042 -	62274\\
              & 62542 -	63076\\
63341 - 64935 & 63466 -	65166\\
              & 65468 -	65933\\
\hline\hline
\end{tabular}
\end{table}
\begin{table}[h!t]
\centering
\caption{\label{Complementary} Sensitivity (TPR), precision (PPV) and number of non-detected VSG (ND-VSG) reached by a HMM with 3 states for the complementary sequences.}
\begin{tabular}{l c c c  c c c  c c c}
  \hline\hline
$\;$  & $\;$ & $A$ & $\;$ & $\;$ & $B$ & $\;$ & $\;$ & $C$ & $\;$\\  
$\;$  & $S_3$ & $S_6$ & $S_9$ & $S_5$ & $S_7$ & $S_8$ & $S_1$ & $S_2$ & $S_4$\\\hline
TPR      & -  & - & 0.9508  & 0.9485 & 0.9568 & 0.9540 & 0.9226 & - & 0.9385 \\
PPV      & - & -  & 0.6157 & 0.5568 & 0.4833 & 0.5422 & 0.5233 & - & 0.6005 \\
ND-VSG &   0  &    0 &  0 &  0   & 0 & 0 & 0 & 0 & 0 \\
\hline\hline
\end{tabular}
\end{table}

\section{Conclusions and Future Work}
\label{Conclusion}
In this article we tackle a classification problem using a \textit{Hidden Markov Model (HMM)}.
%
%
Our approach identified with relatively \textit{success} a particular type of genes on  \textit{Trypanosoma brucei (T. brucei)} genome, those encoding \textit{Variant Surface Glycoproteins (VSG)} from African trypanosomes.
\textit{T. brucei} uses the expression of different VSG proteins to evade host immune mechanisms provoking the Trypanosomiasis disease.
The results shows that the VSG genes have characteristics that make them detectable as statistically homogeneous zones in  \textit{T. brucei} genome.
%
%
%
We perform a HMM with $2$ and $3$ states. The model with $3$ states outperforms the 2-states model, identifying VSG in the two opposite strand directions. 
The HMM with $3$ states is efficient in searching VSG genes, according our experiments in all cases the model reached a sensitivity over the $90\%$ and the totality of VSG genes were well-identified.

In addition, we propose a simple method for identifying the misclassified genes by the HMM model. 
We define a filter using the location of the VSG in the training data.  
Despite the simplicity of the approach presented in this article, according the empirical results for a particular genome the performance for gene detection was very high.
This approach can be of great utility to apply to other African trypanosomes whose silent repertoires of VSG genes are not well-characterized experimentally as well as to others families of hyper-variable genes. 

The present work was restricted to detect VSG genes within regions that are already known to be VSG gene clusters. Therefore the problem was delimited to differentiate only three states, VSG genes, VSG genes located on the complementary strand of DNA and intergenic regions. 
An interesting issue for future work is the following: it would be of great interest to distinguish also between genomic regions containing VSG clusters from other genomic regions containing regular protein coding genes (encoding for normal functions in the cell).







\vskip 0.2in
\bibliography{Biblio}

\end{document}